\documentclass[11pt,a4paper]{article}

\usepackage{amsfonts}
\usepackage[dvips]{graphicx}
\usepackage{amsthm}
\usepackage{epsfig}
\usepackage{t1enc}
\usepackage{amssymb}
\usepackage{latexsym}
\usepackage{bm}

\font\bbx=cmbx10 scaled\magstep2

\title{\bf Perihelion motion \\ due to the Poynting-Robertson
effect}
\author{V. Balek\footnote{Department of Theoretical Physics, Comenius
University, Mlynsk\'a dolina, 842 48 Bratislava, Slovakia; e-mail:
balek@fmph.uniba.sk} \ and J. Kla\v cka\footnote{Astronomical Institute,
Comenius University, Mlynsk\'a dolina, 842 48 Bratislava, Slovakia;
e-mail: klacka@fmph.uniba.sk}}
\begin{document}
\maketitle
\maketitle\abstract

{The force acting on a test particle moving around a body
that is gravitating and emitting electromagnetic radiation 
at the same time is computed up to the second order in $1/c$,
and the shift of perihelion is found. The total shift
is positive (the perihelion moves in the direction of particle 
revolution) and it increases from zero to infinity as the 
radiation pressure to gravity ratio increases from zero to 1.
The additional shift caused by the radiation pressure contains
a term inversely proportional to eccentricity, therefore it 
may dominate over the general relativistic shift if the 
orbit is round enough. 

}


\vskip 5mm
\section{Introduction}
\vskip 1mm
The influence of the radiation pressure on freely moving bodies, 
known as the Poynting-Robertson effect (hereafter PR effect), is 
supposed to play an important role in the dynamics of dust particles
in the Solar System. Since this effect was proposed by Poynting (1903)
and Robertson (1937), radiation-induced corrections to the orbits of 
dust particles were included on a regular basis into astronomical
calculations (for the review, see Gustafson 1994), and the theory 
describing these corrections has been repeatedly extended
(Mediavilla and Buitrago 1989, Kla\v{c}ka 1992, Kla\v{c}ka 2000).
However, one element of the theory is still missing: 
a reliable formula for the contribution of PR effect to the shift 
of perihelion of the orbiting particle. This contribution is of the 
same order of magnitude as the gravitationally induced shift of 
perihelion that 
has been discovered by Einstein and belongs to the three classical 
tests of general relativity; thus if one wants to compute it,
one must first rewrite the theory of PR effect into the general 
relativistic form. This has been done in none of the papers dealing
with the problem so far (Robertson 1937, Wyatt and Whipple 1950, 
Lyttleton 1976). 

\vskip 5mm
In this paper we propose a description of the perihelion motion
due to the PR effect that is based strictly on general relativity
theory. (We use the word 'perihelion' in a broad sense, referring to 
an orbit around a central body that may or may not be the Sun.) 
As a result, we find a formula for the advance of perihelion 
differing from all expressions to be found in the 
literature. Our result can be obtained most easily by solving 
perturbatively the second order equation for the trajectory of the 
particle, however, we follow here a procedure common in astronomical 
community and use an integral formula in which the advance
of perihelion is expressed in terms of the perturbative force.

\vskip 5mm
In section 2 we compute the force to the required
order of perturbation theory, in section 3 we find the shift of
perihelion and in section 4 we discuss the result. Greek indices
run from 0 to 3, latin indices run from 1 to 3, the signature of 
the metric tensor is (+, -, -, -), and the system of units is used 
in which $m = c = r_g = 1$, 
where $m$ is the mass of the test particle and $r_g$ is the 
gravitational radius of the central body given in terms of the mass
of the body $M$ by the standard formula $r_g = 2GM/c^2$. 
Having put $c = 1$ we have not given up the possibility 
to expand the quantities of interest in orders $1/c$: if $v$ is a typical
velocity of the particle and $r$ is a typical radial coordinate,  
the $n$th order in $1/c$ means the $n$th order in $v$ or in $r^{-1/2}$. 
Note, however, that this rule applies only to the quantities 
that are dimensionless in ordinary units; particularly, the Newtonian 
gravitational force, whose magnitude is $1/(2r^2)$ in our units, is
of course of order 1 and not $1/c^4$.


\vskip 1mm
\section{The force up to the order 1/${\bm c}^{\bf 2}$}
\vskip 1mm

Let us first calculate separately the force coming from the PR effect 
in flat spacetime and the general relativistic correction to the 
Newtonian force. Consider a particle in flat spacetime passing by
a pointlike isotropic source of radiation with the luminosity $\cal 
L$. If the distance of the particle from the source is $r$, the 4-velocity
of the particle is $u^\mu$ and the cross section of the particle is $A$,
the 4-vector of force acting on the particle is (Robertson 1937)
\begin{equation}
F_{PR}^\mu = \frac \beta{2r^2} w(l^\mu - wu^\mu),
\label{Fmi} 
\end{equation}
where $\beta$ is the radition pressure to gravity ratio, 
\begin{equation}
\beta = \frac {A\cal L}{2\pi},
\label{beta}
\end{equation}
$l^\mu$ is a 4-component quantity (not a 4-vector) proportional to 
the wave 4-vector $k^\mu = (\omega, {\bm k})$ whose time
component is equal to 1, $l^\mu = k^\mu/\omega$, and $w = u\ .\ l$.
Denote the radius vector of the particle $\bm x$ and the velocity
of the particle $\bm v$. The quantities $u^\mu$, $l^\mu$ and $w$
may be written as $u^\mu = (\gamma, \gamma {\bm v})$, $l^\mu = (1,
{\bm n})$ and $w = \gamma (1 - v_R)$, where $\gamma$ is Lorentz factor,
$\gamma= (1 - v^2)^{-1/2}$,
$\bm n$ is the unit vector in the direction of the vector $\bm x$,
${\bm n} = {\bm x}/r$, and $v_R$ is the radial velocity, $v_R =
{\bm v}\ .\ {\bm n}$. Thus the 3-space part of the 4-force $F_{PR}^\mu$
is
$${\bm F}_{PR} = \frac \beta{2r^2} \gamma (1 - v_R) [{\bm n} -
\gamma^2 (1 - v_R) {\bm v}].$$
If a 4-force is orthogonal to the 4-velocity, $u\ .\ F = 0$, which is
obviously true for $F_{PR}^\mu$, its components may be written
as $F^0 = \gamma \dot \gamma$ and ${\bm F} = \gamma (\gamma{\bm v})\dot{}$,
where the dot denotes the derivative with respect to time, and $F^0$ may
be expressed in terms of $\bm F$ as $F^0 = {\bm v}\ .\ {\bm F}$. Putting
these equations together, one finds after a little algebra that the 
nonrelativistic force (the force of the second Newton law) is
$$ {\bm f} \equiv \dot {\bm v} = \gamma^{-2} [{\bm F} - ({\bm v}\ .
\ {\bm F}) {\bm v}].$$
This expression is valid no matter how many forces act 
on the particle, provided the 4-force we are interested in is 
orthogonal to the 4-velocity. If we insert here the force ${\bm F}_{PR}$
we obtain
$${\bm f}_{PR} = \frac \beta{2r^2} \gamma^{-1} (1 - v_R) ({\bm n} -
{\bm v}),$$
and by expanding this formula in $1/c$ we find that in the zeroth
order the force is equal to $-\beta {\bm f}_N$, where ${\bm f}_N$ is 
the Newtonian force,
\begin{equation}
{\bm f}_N = - \frac {\bm n}{2r^2},
\label{fN}
\end{equation}
and that the full force up to the second order in $1/c$ is
\begin{equation}
{\bm f}_{PR} \doteq -\beta {\bm f}_N + \Delta {\bm f}_{PR}, \ \
\Delta {\bm f}_{PR} \equiv - \frac \beta{2r^2} \left[\left(v_R +
\frac 12 v^2 \right) {\bm n} + (1 - v_R) {\bm v} \right].
\label{fPR} 
\end{equation}

\vskip 5mm
Consider now a nonrelativistic particle moving freely in a weak 
spherically symmetric gravitational field. Our starting point 
will be Schwarzschild metric in isotropic coordinates, 
\begin{equation}
ds^2 \equiv g_{\mu\nu} dx^\mu dx^\nu = F dt^2 - G d{\bm x}^2,
\label{ds2} 
\end{equation}
where $F$ and $G$ are functions of $r$ that are up to the necessary
order in $1/c$ of the form
\begin{equation}
F \doteq 1 - \frac 1r + \frac 1{2r^2}, \ \ G \doteq 1 + \frac
1r.
\label{FG}
\end{equation}
The motion of a free particle is governed by the geodesic equation
\begin{equation}
\frac{Du^\mu}{ds} \equiv \frac{du^\mu}{ds} + \Gamma^\mu_{\nu\kappa}
u^\nu u^\kappa = 0
\label{dumi}
\end{equation}
with the 3-space part
\begin{equation}
\frac{d{\bm u}}{ds} = {\bm F}_g, \ \ F_g^i = - \Gamma^i_{00} (u^0)^2
- 2\Gamma^i_{0j} u^0 u^j - \Gamma^i_{jk} u^j u^k,
\label{du}
\end{equation}
where $\Gamma$'s are coefficients of affine connection defined
in terms of metric tensor in the standard way,
$$\Gamma^{\lambda}_{\mu  \nu} = \frac{1}{2} g^{\lambda \varrho}
\left ( \frac{\partial g_{\varrho \mu}}{\partial x^{\nu}}+
\frac{\partial g_{\varrho \nu}}{\partial x^{\mu}} -
\frac{\partial g_{\mu \nu}}{\partial x^{\varrho}} \right ),
\ \  g^{\mu  \nu} g_{\nu  \varrho} = \delta^{\mu}_{\varrho}.$$
If we insert here the metric tensor $g_{\mu\nu}$ = diag$(F, -G, -G,
-G)$, we find that the nonzero $\Gamma$'s evaluated to the necessary
order in $1/c$ are
$$\Gamma^i_{00} \doteq \frac 1{2r^2} \left(1 - \frac 2r \right) n^i,
\ \ \Gamma^i_{jk} = - \frac 1{2r^2} [\delta_{ij} n^k + \delta_{ik} n^j
- \delta_{jk} n^i],$$
which yields
$${\bm F}_g \doteq - (u^0)^2 \frac 1{2r^2} \left[\left(1 + v^2 - \frac
2r \right) {\bm n} - 2v_R {\bm v} \right].$$
On the other hand, since
$$u^0 \equiv \frac {dt}{ds} = \left(g_{\mu\nu} \frac {dx^\mu}{dt}
\frac {dx^\nu}{dt}\right)^{-1/2} \doteq 1 + \frac 12 v^2 + \frac 1{2r},$$
we have
$$\frac{d{\bm u}}{ds} = u^0 (u^0 {\bm v})\dot{} \doteq
(u^0)^2 \left[\dot{\bm v} + \left(v\dot v - \frac 1{2r^2} v_R \right)
{\bm v}\right] = (u^0)^2 \left[{\bm f} + ({\bm v}\ .\ {\bm f}) {\bm v}
- \frac 1{2r^2} v_R {\bm v}\right].$$
We now identify the force $\bm f$ in the first term in square brackets
with the force ${\bm f}_g$, replace the force $\bm f$ in the second
term in square brackets by the force ${\bm f}_g$ 
in the zeroth approximation, that means by the Newtonian
force (\ref{fN}), and put the expression obtained in this way equal to
the force ${\bm F}_g$ computed earlier. As a result we find
\begin{equation}
{\bm f}_g \doteq {\bm f}_N + \Delta {\bm f}_g, \ \ \Delta {\bm f}_g
\equiv - \frac 1{2r^2} \left[\left(v^2 - \frac 2r\right) {\bm n}
- 4v_R {\bm v}\right].
\label{fg} 
\end{equation}
Note that the same force is obtained from the approximative Lagrangian 
of a system of gravitating particles valid up to the fourth order
in $1/c$ (Landau and Lifshitz 1975).

\vskip 5mm
After these preliminary considerations we are ready to derive the 
expression for the force responsible for PR effect in Schwarzschild 
metric up to the second order of $1/c$. Equation of motion of
the particle is
\begin{equation}
\frac{Du^\mu}{ds} = \Phi_{PR}^\mu,
\label{dumi1} 
\end{equation}
where $\Phi_{PR}^\mu$ is the 4-force responsible for the PR effect 
in curved spacetime. (For later purposes we denote this 4-vector
differently than in flat spacetime, although its physical meaning
is the same.) In flat spacetime, the 4-force responsible 
for the PR effect is in general of the form
$$F_{PR}^\mu = AU w(l^\mu - wu^\mu),$$
where $U$ is the energy density of the radiation.
To see that this expression transforms properly
under Lorentz transformation note that the noninvariant factor
$1/\omega$ appearing in $w$ and $l^\mu$ cancels out
since $U$ is proportional to $\omega^2$. The PR effect involves
the scale of dust particles that is, say, 1 mm, and the atomic scale
entering the considerations if the interaction of radiation with
matter is described by some kind of microscopic theory; and since both
scales are much smaller than the radius of curvature in the neighbouhood
of a body with Sun's mass, which is about 0.1 ly at the Earth-Sun
distance, we may pass from special
to general relativistic formula in the simplest possible way, using
the rule "$\eta_{\mu\nu}$ goes to $g_{\mu\nu}$ and comma goes to 
semicolon". Possible corrections contain the curvature tensor, thus
are of the order of the ratio of a typical scale of the problem to the
radius of curvature. Consequently, the expression for 
$F_{PR}^\mu$ is valid also for $\Phi_{PR}^\mu$,
and one has just to replace Minkowski metric in the definition
of $w$ and in the condition that the 4-vector $k^\mu$ (appearing in 
the definition of $l^\mu$) is a null vector, by the metric of 
curved spacetime. 

\vskip 5mm
In the problem we are interested in the source of radiation is 
a spherically symmetric static body with the mass $M$ radiating 
isotropically with the luminosity (total energy transported trough 
a closed surface in infinity per 1 sec) $\cal L$. 
To obtain the expression for $U$, introduce local inertial observers
that are at the given moment at rest with respect to observers
at infinity.
The time intervals measured by these observers are $\Delta \tau
= F^{1/2} \Delta t$, which implies, first, that $\omega_{LIS}
= F^{-1/2} \omega$ and $U_{LIS} = F^{-1} U$, and second,
that the energy of photons is scaled by a factor $F^{-1/2}$
and time intervals between the photons are scaled by a factor
$F^{1/2}$ with respect to the corresponding quantities measured by
the observers in infinity, thus the total energy ${\cal L}_r$
transported through the sphere with the radius $r$ per 1 sec is
$F^{-1} \cal L$. On the other hand, ${\cal L}_r = U_{LIS} \times$
the area of sphere = $4\pi U_{LIS} G r^2$, hence
$${\cal L} = 4\pi FU_{LIS} G r^2 = 4\pi UG r^2.$$
If one expresses $U$ from here, inserts it into the expression 
of 4-force and uses the definition of $\beta$, one arrives at
\begin{equation}
\Phi_{PR}^\mu = \frac \beta{2Gr^2} w(l^\mu - w u^\mu).
\label{Fmi1} 
\end{equation}
We have obtained an expression that is the same as in flat space,
except for the factor $G$ in the denominator. Note that the 
Schwarzschild radial coordinate equals $G^{1/2}r$ so that in 
Schwarzschild coordinates both expressions coincide.

\vskip 5mm
Now we have to find the 3-space part of the 4-force $\Phi_{PR}^\mu$
up to the second order in $1/c$, and insert the result into the
approximative form of the 3-space part of the equation of motion. 
From the definition of $w$ and $l^\mu$ with the general relativistic 
scalar product $a\ .\ b = g_{\mu\nu} a^\mu b^\nu$, using 
the metric tensor of equation (\ref{ds2}) we obtain
$l^\mu = (1, (F/G)^{1/2} {\bm n})$ and $w = u^0 [F - (FG)^{1/2}v_R]$,
so that ${\bm l} \doteq (1 - 1/r) {\bm n}$, $w \doteq u^0 (1 -
v_R - 1/r) \doteq (u^0)^2 [1 - v_R - v^2/2 - 3/(2r)]$, and
$${\bm \Phi}_{PR} = \frac \beta{2Gr^2} w \left({\bm l}
-u^0 w {\bm v}
\right) \doteq (u^0)^2 \frac \beta{2r^2} \left[\left(1 - v_R - 
\frac 12 v^2 - \frac 7{2r} \right) {\bm n} - (1 - 2v_R) {\bm v}
\right].$$
The 3-space part of the equation of motion is
\begin{equation}
\frac{d{\bm u}}{ds} = {\bm F}_g + {\bm \Phi}_{PR},
\label{du1}
\end{equation}
and if we express the left hand side in the same way as earlier, 
we find
$${\bm f} \doteq {\bm f}_g + (u^0)^{-2} {\bm \Phi}_{PR} -
({\bm v}\ .\ {\bm f}_{PR}^{(0)}) {\bm v},$$
where ${\bm f}_{PR}^{(0)}$ is the force responsible for the
PR effect in the zeroth approximation, ${\bm f}_{PR}^{(0)}
= -\beta {\bm f}_N$. Finally, inserting here the approximative expression
for ${\bm \Phi}_{PR}$ and denoting $b = 1 -\beta$, we obtain
\begin{equation}
{\bm f} \doteq b {\bm f}_N + \Delta {\bm f}_g + \Delta {\bm f}_{PR} +
\delta {\bm f}, \ \ \delta {\bm f} = - \frac {7\beta}{4r^3} {\bm n}.
\label{f1}
\end{equation}
Thus the total force consists of a modified Newtonian force
$b {\bm f}_N$ corresponding to the effective mass of the source 
$M_{eff} = bM$, the correction to the gravitational force 
$\Delta {\bm f}_g$ given in equation (\ref{fg}), the correction to 
the force responsible for PR effect $\Delta {\bm f}_{PR}$
given in equation (\ref{fPR}), and an additional force $\delta {\bm f}$.


\vskip 1mm
\section{Perihelion motion}
\vskip 1mm
Consider a particle moving on a bounded orbit under the combined 
action of Newtonian gravitational force of a pointlike source with 
the mass $M_{eff}$ and a small perturbing force acting in the plane
of motion. In the first order in perturbing force, the perihelion
shifts during one period by the angle 
\begin{equation}
\Delta \phi = \int_0^{2\pi} \frac 2{be} \left(-R c + Ts \frac{2 +
ec}{1 + ec}\right) r^2 d\phi,
\label{dfi} 
\end{equation}
where $R$ and $T$ are the radial and transversal components of the 
perturbative force, $c = \cos \phi$, $s = \sin \phi$ and $e$ is the
eccentricity of the unperturbed orbit (an ellipse with one focus placed
at the source). After inserting the expressions for $R$ and $T$
one has to replace the functions $r$, $v_R$ and $v_T$ (the transversal
component of velocity) by the limit form they assume for the unperturbed
orbit,
\begin{equation}
r = \frac p{1 + ec}, \ \ v_R = \sqrt{\frac b{2p}}es, \ \ v_T = 
\sqrt{\frac b{2p}} (1 + ec),
\label{rv} 
\end{equation}
where $p$ is the parameter of the unperturbed orbit. The parameter and 
the eccentricity of the unperturbed orbit are defined in terms
of the energy $E$ and the angular momentum $L$ of the particle as
\begin{equation}
p = \frac 2b L^2, \ \ e = \sqrt{1 + \frac 8{b^2} EL^2}.
\label{pe} 
\end{equation} 

\vskip 5mm
A simple argument leading to the formula (\ref{dfi}) is given in Burns 
(1976). Let us summarize this argument in a slightly modified form, with
the time dependence of all quantities replaced by the dependence on the
angular coordinate. One starts with the Ansatz that the particle
is placed at the {\it osculating orbit}
\begin{equation}
r = \frac {p(\phi)}{1 + e(\phi) \hat c},
\label{r}
\end{equation}
where $\hat c = \cos [\phi - \omega(\phi)]$ with the function $\omega
(\phi)$ to be determined, and $p(\phi)$ and $e(\phi)$
are so called {\it osculating elements} given by relations (\ref{pe})
with the constant energy and angular momentum of the unperturbed orbit
replaced by the variable energy and angular momentum of
the perturbed orbit,
$E \to E(\phi)$ and $L \to L(\phi)$. To determine $\omega$ one
postulates that an 'incomplete total derivative' of $r$ with respect to
$\phi$, defined as the sum of all terms entering the total derivative
except for the partial derivative with respect to $\phi$ at given $p$,
$e$ and $\omega$, is zero,
\begin{equation}
\frac {d'r}{d\phi} \equiv \frac {\partial r}{\partial p}
\ \frac {dp}{d\phi} + \frac {\partial r}{\partial e} \ \frac
{de}{d\phi} + \frac {\partial r}{\partial \omega} \ \frac {d\omega}
{d\phi} = 0.
\label{dr}
\end{equation}
If one puts $\Delta \phi = \omega (2\pi) - \omega (0)$,
one can verify by a straightforward computation that this Ansatz gives
in the first order in perturbing force the right equation for $\Delta 
\phi$. This is of
course not an accident. In fact, from the definition of energy and angular
momentum it can be seen that the Ansatz (\ref{r}) and (\ref{dr}) provides
the exact solution to the problem. Consequently, the expression
of the angle $\Delta \phi$ in the form (\ref{dfi}), with the functions
$r$, $v_R$ and $v_T$ given in (\ref{rv}), is valid to all orders of
perturbation theory if one regards the quantities $p$ and $e$ as
functions of $\phi$ and replaces $c \to \hat c$ and $s \to \hat s
\equiv \sin [\phi - \omega(\phi)]$.

\vskip 5mm
An appropriate reference value for the angle $\Delta \phi$ is the
general relativistic shift of perihelion in the field generated by 
a source with the mass $M_{eff}$,
\begin{equation}
\Delta \phi_E = \frac {3\pi b}{2L^2}
\label{dfE}
\end{equation}
('E' stands for 'Einstein'). If we express the angular momentum $L$
in terms of the major semiaxis $a$ and the eccentricity $e$ and pass to
the ordinary units, this reduces to the standard formula
\begin{equation}
\Delta \phi_E = \frac {3\pi r_g}{a(1 - e^2)}.
\label{dfE1}
\end{equation}
To obtain $\Delta \phi$ for perturbative forces of the type
$v^2/r^2$ or $1/r^2$ it is sufficient to do calculations in the
first order of perturbation theory. If the force is of the form
$$\Delta {\bm f} = \frac 1{r^2} \left(Av^2 {\bm n} + Bv_R {\bm v}
+ C \frac {\bm n}r \right)$$
we obtain
\begin{equation}
\Delta \phi \doteq -\frac 23 \left(A - B + \frac 1b C\right).
\label{dfi1}
\end{equation}
On the other hand, for the force of the form $v/r^2$ the second order
of the perturbation theory is needed. If the force is of the form
$$\Delta {\bm f} = \frac 1{r^2} \left({\cal A}v_R {\bm n} + {\cal B}
{\bm v}\right),$$
using the relations
$$\dot E = v_R R + v_T T, \ \ \dot L = rT, \ \ r^2 \dot
\phi = L,$$
and the expression of $e$ in terms of $E$ and $L$ we find
$$\frac {dL}{d\phi} = {\cal B}, \ \ \frac {de}{d\phi}
\doteq \frac 1L [2{\cal B}c + e({\cal A}s^2 + 2{\cal B})],$$
hence
$$\Delta L = {\cal B} \phi, \ \ \Delta e \doteq
\frac 1L \left[2{\cal B}s + e\left(\frac 12 {\cal A} + 2{\cal B}
\right)\phi - \frac 14 e {\cal A} S \right],$$
where $S = \sin (2\phi)$. The derivative of $\omega$ is approximately
equal to the integrand of (\ref{dfi}), which in our case leads to
$$\frac {d\omega}{d\phi} \doteq - \frac 1L \left({\cal A} c -
\frac 2{e} {\cal B} \right) s,$$
thus the increment of $\omega$ is
$$\Delta \omega \doteq \frac 1L \left[\frac 14{\cal A} (C - 1) -
\frac 2{e} {\cal B} (c - 1) \right],$$
where $C = \cos (2\phi)$. Now we insert these increments into
\begin{eqnarray*} 
\lefteqn {\Delta \phi = - \int_0^{2\pi} \frac 1{L(\phi)} \left[{\cal A}
\hat c - \frac 2{e(\phi)} {\cal B} \right] \hat s d\phi} \\ & & \doteq
\frac 1L \int_0^{2\pi} \left[\left({\cal A} c - \frac 2e {\cal B}
\right) s \frac {\Delta L}L \right.  
\left. - \frac 2{e^2} {\cal B} s \Delta e +
\left({\cal A} C - \frac 2e {\cal B} c\right) \Delta \omega\right]
d\phi, \end{eqnarray*} 
and we find
\begin{equation}
\Delta \phi \doteq \frac 8{3b} \left[\frac 1{16} {\cal A}\left({\cal A}
- 2{\cal B}\right) + \frac 1e {\cal B}\left(\frac 12{\cal A} + 3{\cal B}
\right)\right] \Delta \phi_E.
\label{dfi2}
\end{equation}
Using (\ref{dfi1}) and (\ref{dfi2}) we obtain shifts of perihelion
due to the three parts of the perturbative force in (\ref{f1})
$$\Delta \phi_g = \left(\frac 53 - \frac 2{3b}\right) \Delta \phi_E,
\ \ \Delta \phi_{PR} = \beta \left[\frac 12 + \frac \beta{3b}
\left(-\frac 18 + \frac 7e \right)\right] \Delta \phi_E, \ \
\delta \phi = \frac {7\beta}{6b} \Delta \phi_E,$$
and summing them up we find the total shift of perihelion
\begin{equation}
\Delta \phi = \frac 1b \left[1 + \frac 13 \beta^2
\left(-\frac {13}8 + \frac 7e \right)\right] \Delta \phi_E.
\label{dfi0}
\end{equation}


\vskip 1mm
\section{Conclusion}
\vskip 1mm
As we have seen, when describing the perihelion motion due to the
PR effect one has to perform the expansion to the required order
of $1/c$ in the general relativistic formulation 
of the full theory, rather than put together the effects of gravity
and radiation pressure computed to the required order in general 
relativity without radiation, and in Maxwell theory without gravity. 
The additional force we have obtained may be regarded
as a sort of interference term describing the mixing between
the effects of gravity and radiation pressure. This is clear from 
the very fact that this force belongs neither to the purely
gravitational not to the purely radiational part of the total force,  
but we can see the point also explicitly if we insert into the
expression of the additional force from the definition of $\beta$
and rewrite
the resulting expression into ordinary units. We arrive at
$$\delta {\bm f} = - \frac{7GMA{\cal L}}{4\pi c^3 r^3} {\bm n},$$
so that the additional force depends on both characteristics of the 
central body relevant 
for the problem: on the mass $M$ characterizing the body as the source
of gravitational force, as well as on the luminosity $\cal L$
characterizing the body as the source radiation.

\vskip 5mm
The expression for the advance of perihelion is inversely 
proportional to $b \equiv 1 - \beta$, thus it diverges as $\beta$ 
approaches 1. Of course, our formula cannot be used for values of
$\beta$ arbitrarily close to this limit. The perturbation
theory breaks down when the dominant term in the perturbative force, 
linear in velocity and independent on $b$, becomes of the same 
order of magnitude as the zeroth order force proportional to $b$.
This leads, in ordinary units, to the condition $b \gg v/c$,
which means that if a particle is orbiting around the Sun with Earth's
velocity, its parameter $\beta$ must differ from 1 by a value much
greater than $10^{-4}$.

\vskip 5mm
A subtle point of the computation is that one has to evaluate
the contribution of the perturbative force linear in
velocity in the next-to-leading order. The net effect is that a term
proportional to $1/e$ appears in the result, which means that the shift
of perihelion caused by the PR effect dominates over the general 
relativistic shift for orbits that are sufficiently round. Again,
there is a limit to the applicability of our formula. The calculation 
within the perturbation theory 
is no longer possible if the term in question assumes a value of lower 
order in $1/c$ than other terms coming from the PR effect, that means 
if the expression $\beta^2/e$ becomes comparable with $\beta r^{1/2}$. 
This leads to the condition $e \gg \beta r^{-1/2}$, or in ordinary 
units 
$$e \gg \beta \sqrt{\frac {r_g}r}.$$
If a particle has the size of order 1 mm, which means that its
$\beta$ is of order $10^{-3}$, and if it revolves around the Sun at
the distance of Earth, our formula is valid for eccentricities
ranging from 1 down to the values of order $10^{-7}$, becoming a rough
estimate at the lower limit. The behaviour of orbits with even lower 
eccentricity has been investigated by Kla\v{c}ka and Kaufmannov\'{a} (1992).

\vskip 5mm
\noindent
{\it Acknowledgements.} This work was supported by the grants VEGA 1/7069/20
and VEGA 1/7067/20.

\vskip 5mm
\noindent
{\bbx References}
\vskip 5mm
\noindent
Burns J. A. 1976. Elementary Derivation of the Perturbation
Equations of Celestial Mechanics. {\it Am. J. Phys.} {\bf 44}, 
944-949.
\vskip 5mm
\noindent
Gustafson B. A. S. 1994. Physics of zodiacal dust.
{\it Annu. Rev. Earth Planet. Sci.} {\bf 22}, 553-595.
\vskip 5mm
\noindent
Kla\v{c}ka J. 1992. Poynting-Robertson effect. I. Equation of
motion.
{\it Earth, Moon, and Planets} {\bf 59}, 41-59.
\vskip 5mm
\noindent
Kla\v{c}ka J. 2000. Electromagnetic radiation and motion of real
particle, astro-ph/0008510.
\vskip 5mm
\noindent
Kla\v{c}ka J., Kaufmannov\'{a} J. 1992. Poynting-Robertson effect:
`circular' orbit. {\it Earth, Moon, and Planets} {\bf 59}, 97-102.
\vskip 5mm
\noindent
Landau L. D., Lifshitz E. M. 1975. {\it The Classical Theory of Fields}.
4th ed., Pergamon Press, New York. (In Russian: {\it Teoria polia}. Nauka,
Moscow, 1973.)
\vskip 5mm
\noindent
Lyttleton R. A. 1976. Effects of solar radiation on the orbits
of small particles. {\it Astrophys. Space Sci.} {\bf 44}, 119-140.
\vskip 5mm
\noindent
Mediavilla E., Buitrago J. 1989. The Poynting-Robertson effect
and its generalization to the case of an extended source in rotation.
{\it Eur. J. Phys.}, {\bf 10}, 127-132. 
\vskip 5mm
\noindent
Poynting J. H. 1903. Radiation in the solar system, its effect on
temperature and its pressure on small bodies.
{\it Phil. Trans. Roy. Soc.} {\bf 202}, 525.
\vskip 5mm
\noindent
Robertson H. P. 1937. The dynamical effects of radiation in the
Solar System. {\it Mon. Not. R. Astron. Soc.} {\bf 97}, 423-438.
\vskip 5mm
\noindent
Wyatt S. P., Whipple F. L. 1950. The Poynting-Robertson effect on
meteor orbits. {\it Astrophys. J.} {\bf 111}, 558-565.

\end{document}